\documentclass[12pt]{article}
\usepackage{lscape}
\usepackage{citesort}
\usepackage{amssymb}
\usepackage{appendix}
\usepackage{multirow}

\begin{document}
\def\qq{\langle \bar q q \rangle}
\def\uu{\langle \bar u u \rangle}
\def\dd{\langle \bar d d \rangle}
\def\sp{\langle \bar s s \rangle}
\def\GG{\langle g_s^2 G^2 \rangle}
\def\Tr{\mbox{Tr}}
\def\figt#1#2#3{
        \begin{figure}
        $\left. \right.$
        \vspace*{-2cm}
        \begin{center}
        \includegraphics[width=10cm]{#1}
        \end{center}
        \vspace*{-0.2cm}
        \caption{#3}
        \label{#2}
        \end{figure}
	}
	
\def\figb#1#2#3{
        \begin{figure}
        $\left. \right.$
        \vspace*{-1cm}
        \begin{center}
        \includegraphics[width=10cm]{#1}
        \end{center}
        \vspace*{-0.2cm}
        \caption{#3}
        \label{#2}
        \end{figure}
                }

\def\ds{\displaystyle}
\def\beq{\begin{equation}}
\def\eeq{\end{equation}}
\def\bea{\begin{eqnarray}}
\def\eea{\end{eqnarray}}
\def\beeq{\begin{eqnarray}}
\def\eeeq{\end{eqnarray}}
\def\ve{\vert}
\def\vel{\left|}
\def\ver{\right|}
\def\nnb{\nonumber}
\def\ga{\left(}
\def\dr{\right)}
\def\aga{\left\{}
\def\adr{\right\}}
\def\lla{\left<}
\def\rra{\right>}
\def\rar{\rightarrow}
\def\lrar{\leftrightarrow}  
\def\nnb{\nonumber}
\def\la{\langle}
\def\ra{\rangle}
\def\ba{\begin{array}}
\def\ea{\end{array}}
\def\tr{\mbox{Tr}}
\def\ssp{{\Sigma^{*+}}}
\def\sso{{\Sigma^{*0}}}
\def\ssm{{\Sigma^{*-}}}
\def\xis0{{\Xi^{*0}}}
\def\xism{{\Xi^{*-}}}
\def\qs{\la \bar s s \ra}
\def\qu{\la \bar u u \ra}
\def\qd{\la \bar d d \ra}
\def\qq{\la \bar q q \ra}
\def\gGgG{\la g^2 G^2 \ra}
\def\GG{\langle g_s^2 G^2 \rangle}
\def\g5{\gamma_5 \not\!q}
\def\x{\gamma_5 \not\!x}
\def\g5{\gamma_5}
\def\sb{S_Q^{cf}}
\def\sd{S_d^{be}}
\def\su{S_u^{ad}}
\def\sbp{{S}_Q^{'cf}}
\def\sdp{{S}_d^{'be}}
\def\sup{{S}_u^{'ad}}
\def\ssp{{S}_s^{'??}}

\def\sig{\sigma_{\mu \nu} \gamma_5 p^\mu q^\nu}
\def\fo{f_0(\frac{s_0}{M^2})}
\def\ffi{f_1(\frac{s_0}{M^2})}
\def\fii{f_2(\frac{s_0}{M^2})}
\def\O{{\cal O}}
\def\sl{{\Sigma^0 \Lambda}}
\def\es{\!\!\! &=& \!\!\!}
\def\ap{\!\!\! &\approx& \!\!\!}
\def\ar{&+& \!\!\!}
\def\arrr{\!\!\!\! &+& \!\!\!}
\def\ek{&-& \!\!\!}
\def\vev{&\vert& \!\!\!}
\def\kek{\!\!\!\!&-& \!\!\!}
\def\cp{&\times& \!\!\!}
\def\se{\!\!\! &\simeq& \!\!\!}
\def\eqv{&\equiv& \!\!\!}
\def\kpm{&\pm& \!\!\!}
\def\kmp{&\mp& \!\!\!}
\def\mcdot{\!\cdot\!}
\def\erar{&\rightarrow&}


\def\simlt{\stackrel{<}{{}_\sim}}
\def\simgt{\stackrel{>}{{}_\sim}}


\title{
         {\Large
                 {\bf
Magnetic moments of negative parity heavy baryons in QCD
                 }
         }
      }

\author{\vspace{1cm}\\
{\small T. M. Aliev$^1$ \thanks {e-mail: taliev@metu.edu.tr}}\,\,
{\small K. Azizi$^2$    \thanks {e-mail: kazizi@dogus.edu.tr}}\,\,,
{\small M. Savc{\i}$^1$ \thanks {e-mail: savci@metu.edu.tr}} \\
{\small $^1$ Department of Physics, Middle East Technical University,
06531 Ankara, Turkey}\\
{\small $^2$ Department of Physics, Do\u gu\c s University,
Ac{\i}badem-Kad{\i}k\"oy, 34722 \.{I}stanbul, Turkey}}

\date{}

\begin{titlepage}
\maketitle
\thispagestyle{empty}

\begin{abstract}

The magnetic moments of the negative parity, spin-1/2 baryons containing
single heavy quark are calculated.
The pollution that occur
from the transitions between positive and negative parity baryons are
removed by constructing the sum rules from different Lorentz structures.

\end{abstract}

~~~PACS numbers: 11.55.Hx, 13.40.Em, 14.20.Lq, 14.20.Mr

\end{titlepage}

\section{Introduction}

Last decade was quite fruitful in the field of heavy baryon spectroscopy.
At present, all baryons containing heavy charm and bottom quarks, except
$\Omega_b^\ast$ baryon, as well as several heavy baryons with negative
parity, have been observed by a number of collaborations (for a review see
\cite{Rsbs01}). These progresses stimulated further theoretical studies, and
experimental researches on heavy baryons at the existing facilities,
especially at LHC.

Heavy baryons are well recognized to represent a rich ``laboratory" for
theoretical investigations. The properties of heavy baryons have been
studied in framework of the various methods such as, relativistic quark
model \cite{Rsbs02}, variational approach \cite{Rsbs03}, constituent
quark model \cite{Rsbs04}, lattice QCD model \cite{Rsbs05}, QCD sum
rules method \cite{Rsbs06}. Recent progress on this subject can be found in
\cite{Rsbs07}.

One of the most crucial quantities in discovering the internal structure of
the baryons is their magnetic moments. Magnetic moments of the ground
state $J^P = {1\over 2}^+$ heavy baryons have widely been studied
in literature. Furthermore, magnetic moments of the heavy baryons have been
investigated in the naive quark model in \cite{Rsbs08,Rsbs09},
phenomenological quark model \cite{Rsbs10}, relativistic three-quark model
\cite{Rsbs11}, variational approach \cite{Rsbs12}, nonrelativistic quark
model with screening and effective quark mass \cite{Rsbs13,Rsbs14},
nonrelativistic hypercentral model\cite{Rsbs15}, chiral constituent
quark model \cite{Rsbs16}, chiral bag model \cite{Rsbs17}, chiral
perturbation theory \cite{Rsbs18}, traditional QCD sum rules \cite{Rsbs19},
and light cone QCD sum rules (LCSR)
\cite{Rsbs20,Rsbs21,Rsbs22,Rsbs23}, respectively.

In the present work, we calculate magnetic moments of the negative parity,
spin-1/2 heavy baryons in framework of the light cone QCD sum rules method
(for a review about this method, see for example \cite{Rsbs24}).

The body of the paper is organized as follows. In section 2, we construct
the light cone QCD sum rules for the negative parity, heavy baryons.
Section 3 is devoted to the numerical analysis of these sum rules
for the magnetic moments. This section also contains discussion of the
obtained results, and comparison with the prediction of the other
approaches.   

\section{Construction of the sum rules for magnetic moments of the negative
parity heavy baryons}

In this section the LCSR for the negative parity baryons are derived. In
order to formulate the relevant sum rules, we consider the following
correlator,
\bea
\label{esbs01}
\Pi (p,q) = i \int d^4x e^{ipx} \lla 0 \vel \mbox{\rm T} \{
\eta_Q (x) \bar{\eta}_Q(0) \}\ver 0 \rra_\gamma ~,
\eea
where $\gamma$ means the external magnetic field, $\eta(x)$ is the
interpolating current of the corresponding heavy baryon with spin-1/2. 
In order to obtain the sum rules for magnetic moments of the heavy baryons
the correlation function function is calculated in two different ways:
a) In terms of the hadrons; b) performing the operator product expansion (OPE) over
the twists of operators, and using the photon distribution amplitudes (DAs) which
encode all nonperturbative effects. The sum rules for the magnetic moments
of the heavy baryons are obtained by equating the coefficients of the
appropriate Lorentz structures that survive in parts (a) and (b). Finally,
in order to enhance the contributions coming from the ground states, and
suppress the contributions of the continuum and higher states, Borel
transformation with respect to the momentum squared of the initial and final
baryon states, and continuum subtraction procedure are performed,
successively (for more about the LCSR, see for example \cite{Rsbs24}).

To be able to calculate the correlator function in terms of the hadrons, we
insert the complete set of baryon states that carry the same quantum numbers
as the interpolating current $\eta(x)$. It should be noted here that the
interpolating current can produce both positive and negative parity baryons
from the vacuum state. Keeping this remark in mind, and isolating the
contributions of the ground state baryons, we get
\bea
\label{esbs02}
\Pi(p,q) \es
{\la 0 \ve \eta \ve f_+(p,s) \ra \over p^2-m_{f+}^2} \la f_+(p,s) \ve
i_+(p+q,s) \ra_\gamma {\la i_+(p+q,s) \ve \bar{\eta}(0) \ra \over
(p+q)^2-m_{i+}^2} \nnb \\
\ar {\la 0 \ve \eta \ve f_-(p,s) \ra \over p^2-m_{f-}^2} \la f_-(p,s) \ve
i_+(p+q,s) \ra_\gamma {\la i_+(p+q,s) \ve \bar{\eta}(0) \ra \over
(p+q)^2-m_{i+}^2} \nnb \\
\ar {\la 0 \ve \eta \ve f_+(p,s) \ra \over p^2-m_{f+}^2} \la f_+(p,s) \ve
i_-(p+q,s) \ra_\gamma {\la i_-(p+q,s) \ve \bar{\eta}(0) \ra \over
(p+q)^2-m_{i-}^2} \nnb \\
\ar {\la 0 \ve \eta \ve f_-(p,s) \ra \over p^2-m_{f-}^2} \la f_-(p,s) \ve
i_-(p+q,s) \ra_\gamma {\la i_-(p+q,s) \ve \bar{\eta}(0) \ra \over
(p+q)^2-m_{i-}^2} + \cdots~,
\eea
where $f_{+(-)}$ and $i_{+(-)}$ correspond to the final and initial,
positive (negative) parity baryonic states with spin $s$; and
$m_{f+(-)}$ and $m_{i+(-)}$ to their masses, respectively; $q$ is the
four-momentum of the photon; $\cdots$ describe the contributions coming from
higher states. In this expression, the first term describes
positive to positive, the second term describes positive to negative, the
third term describes negative to positive, and the fourth term describes
negative to negative parity transitions, respectively. 

The matrix elements in Eq. (\ref{esbs02}) are determined as,
\bea
\label{esbs03}
\la 0 \ve \eta \ve +(p)\ra \es \lambda_+ u_+ (p)~,\nnb \\
\la 0 \ve \eta \ve -(p)\ra \es \lambda_- \gamma_5 u_- (p)~,\nnb \\
\la +(p) \ve \eta \ve +(p+q)\ra_\gamma \es e \varepsilon^\mu \bar{u}(p)
\left[\gamma_\mu f_1 - {i \sigma_{\mu\nu} q^\nu \over 2 m_+} f_2 \right]
u(p+q)~, \nnb \\
\la -(p) \ve \eta \ve +(p+q)\ra_\gamma \es e \varepsilon^\mu \bar{u}(p)
\left[\gamma_\mu f_1^T - {i \sigma_{\mu\nu} q^\nu \over m_+ + m_-} f_2^T
\right] \gamma_5 u(p+q)~, \nnb \\
\la -(p) \ve \eta \ve -(p+q)\ra_\gamma \es e \varepsilon^\mu \bar{u}(p)
\left[\gamma_\mu f_1^\ast - {i \sigma_{\mu\nu} q^\nu \over 2 m_-} f_2^\ast
\right] u(p+q)~,
\eea
where $\varepsilon^\mu$ and $q^\mu$ are the four-momentum and -polarization
vectors of a photon. Using the Gordon identity for the diagonal transitions,
it can easily be shown that in the case a real photon is exchanged, i.e.
$q^2=0$, the structure $\gamma_\mu$ is proportional to
$f_1+f_2~(f_1^\ast+f_2^\ast)$, and it describes the magnetic moment of the
corresponding baryon. As has already been noted, the magnetic moments of the
heavy $J^P={1\over 2}^+$ baryons are calculated in \cite{Rsbs20,Rsbs23},
and hence, in the present work we calculate the magnetic moments of the
$J^P={1\over 2}^-$ baryons in frame work of the LCSR.

Using the definition of the matrix elements given in Eq. (\ref{esbs03}), and
performing summation over the spins of the heavy baryons, the correlation
function can be written as,
\bea
\label{esbs04} 
\Pi (p,q) \es{\lambda_+^2 (f_1+f_2) \over (m_+^2-p_2^2)(m_+^2-p_1^2)}
({\not\!{p}}_2 + m_+) 
\not\!{\varepsilon} ({\not\!{p}}_1 + m_+)~\nnb \\
\ar {\lambda_-^2 (f_1^\ast+f_2^\ast) \over (m_-^2-p_2^2)(m_-^2-p_1^2)}
({\not\!{p}}_2 - m_-)       
\not\!{\varepsilon} ({\not\!{p}}_1 - m_-)~\nnb \\
\ar {\lambda_+ \lambda_- \over (m_-^2-p_2^2)(m_+^2-p_1^2)}
\left[f_1^T + {m_- - m_+ \over  m_- + m_+} f_2^T \right]
({\not\!{p}}_2 - m_-)       
\not\!{\varepsilon} ({\not\!{p}}_1 + m_+)~\nnb \\
\ar {\lambda_- \lambda_+ \over (m_+^2-p_2^2)(m_-^2-p_1^2)}          
\left[f_1^T + {m_- - m_+ \over  m_- + m_+} f_2^T \right]
({\not\!{p}}_2 + m_+)       
\not\!{\varepsilon} ({\not\!{p}}_1 - m_-)~,
\eea
Denoting
\bea
\label{esbs05}
A \es {\lambda_+^2 (f_1+f_2) \over (m_+^2-p_2^2)(m_+^2-p_1^2)}~,\nnb \\
B \es {\lambda_-^2 (f_1^\ast+f_2^\ast) \over
(m_-^2-p_2^2)(m_-^2-p_1^2)}~,\nnb \\
C \es {\lambda_+ \lambda_- \over (m_-^2-p_2^2)(m_+^2-p_1^2)}
\left[f_1^T + {m_- - m_+ \over  m_- + m_+} f_2^T \right]~,\nnb \\
D \es {\lambda_- \lambda_+ \over (m_+^2-p_2^2)(m_-^2-p_1^2)}
\left[f_1^T + {m_- - m_+ \over  m_- + m_+} f_2^T \right]~,
\eea
the phenomenological part of the correlation function is given as,
\bea
\label{nolabel99}
&& A ({\not\!{p}}_2 + m_+)\not\!{\varepsilon} ({\not\!{p}}_1 + m_+) +
B ({\not\!{p}}_2 - m_-)\not\!{\varepsilon} ({\not\!{p}}_1 - m_-) \nnb \\
\ar C ({\not\!{p}}_2 - m_-)\not\!{\varepsilon} ({\not\!{p}}_1 + m_+) +
D ({\not\!{p}}_2 + m_+)\not\!{\varepsilon} ({\not\!{p}}_1 - m_-)~. \nnb
\eea
This expression contains four type, positive--positive, negative--negative,
positive--negative, negative --positive transitions. Among all transitions
only the coefficient $B$ contains negative--negative parity transition,
whose solution at $q^2=0$ is $(f_1^\ast+f_2^\ast)\ve_{q^2=0}$ corresponds to
the magnetic moment of the negative parity heavy baryons. Having four
unknowns, we choose
the four structures $(\varepsilon\!\cdot\! p) I$,
$(\varepsilon\!\cdot\! p) \rlap/{p}$, $\rlap/{\varepsilon} \rlap/{p}$ and
$\rlap/{\varepsilon}$ which each leads to its own equation. Solving this
system of four linear equations for the above-mentioned unknowns, we can
easily find the unknown variable $B$.

In order to obtain the sum rules we have to calculate the invariant
functions $\Pi_i$ from the theoretical side, for which the
interpolating current of the corresponding hadrons must be known. According
to the $SU(3)_f$ classification, hadrons with single heavy quark belong
to either sextet-symmetric or antitriplet-antisymmetric
flavor representations. Therefore
interpolating currents of the sextet (antitriplet) representation should be
symmetric (antisymmetric) with respect to the light flavors. The heavy
baryons $\Sigma_Q$, $\Xi_Q^\prime$ and $\Omega_Q$ belong to the sextet,
$\Xi_Q$ and $\Lambda_Q$ belong to the antitriplet representations of the
$SU(3)_f$ group.

Using this fact, the general form of the interpolating currents belonging to
the sextet or antitriplet representation of the $SU(3)_f$ group can be
written in the following form (see for example \cite{Rsbs25}),
\bea
\label{esbs06}
\eta^{(s)} \es -{1\over \sqrt{2}} \varepsilon^{abc} \Big\{
(q_1^{aT} C Q^b) \gamma_5 q_2^c +
t (q_1^{aT} C \gamma_5 Q^b) q_2^c +
(q_2^{aT} C Q^b) \gamma_5 q_1^c + (q_2^{aT} C
\gamma_5 Q^b) q_1^c\Big\}~, \nnb \\ 
\eta^{(a)} \es -{1\over \sqrt{6}} \varepsilon^{abc} \Big\{
2 (q_1^{aT} C q_2^b) \gamma_5 Q^c +
2 t (q_1^{aT} C \gamma_5 q_2^b) Q^c +
(q_1^{aT} C Q^b) \gamma_5 q_2^c \nnb \\
\ar t (q_1^{aT} C \gamma_5 Q^b) q_2^c -
(q_2^{aT} C Q^b) \gamma_5 q_1^c -         
t (q_2^{aT} C \gamma_5 Q^b) q_1^c\Big\}~,
\eea
where $t$ is an arbitrary parameter ($t=-1$ corresponds to the
so-called Ioffe current); $a,b,c$ are the color indices; and $C$ is the
charge conjugation operator.
The light quark contents of the sextet and antitriplet representations are
given in Table 1.     


\begin{table}[h]

\renewcommand{\arraystretch}{1.3}
\addtolength{\arraycolsep}{-0.5pt}
\small
$$
\begin{array}{|c|c|c|c|c|c|c|c|c|c|c|}
\hline \hline   
 & \Sigma_{c(b)}^{(++)+} & \Sigma_{c(b)}^{+(0)} & \Sigma_{c(b)}^{0(-)} &
             \Xi_{c(b)}^{\prime 0(-)}  & \Xi_{c(b)}^{\prime +(0)}  &
 \Omega_{c(b)}^{0(-)} & \Lambda_{c(b)}^{+(0)} & 
\Xi_{c(b)}^{0(-)}  & \Xi_{c(b)}^{+(0)} \\   
\hline \hline
q_1   & u & u & d & d & u & s & u & d & u \\
q_2   & u & d & d & s & s & s & d & s & s \\
\hline \hline 
\end{array}
$$
\caption{Quark contents of the heavy baryons belonging to the
sextet and antitriplet representations.}
\renewcommand{\arraystretch}{1}
\addtolength{\arraycolsep}{-1.0pt}
\end{table}      


Using the interpolating currents given in Eq. (\ref{esbs06}),
one can easily calculate the theoretical part of the
correlator function. As an example, we present the form of the correlation
function for $\Sigma_Q^+$ in terms of the corresponding propagators,
\bea
\label{esbs08}
\Pi^{\Sigma_Q^+} \es
-2 \varepsilon^{abc} \varepsilon^{a^\prime b^\prime c^\prime} \int d^4x \la
0 \ve \sum_{\ell=1}^2 \sum_{k=1}^2 \Big\{
A_2^\ell S_u^{cc^\prime} (x) A_2^k \mbox{Tr} S_Q^{bb^\prime} (x) C A_1^k
S_u^{aa^\prime} (x) C A_1^\ell \nnb \\
\ar A_2^\ell \Big[ S_u^{cc^\prime} (x) (C A_1^k)^T S_Q^{bb^\prime T} (x)
(C A_1^\ell)^T S_u^{aa^\prime} (x) A_2^k \nnb \\
\ar S_u^{aa^\prime} (x) (C A_1^k)^T S_Q^{bb^\prime T} (x)
(C A_1^\ell)^T S_u^{cc^\prime} (x) A_2^k \nnb \\
\ar S_u^{aa^\prime} (x) A_2^k \mbox{Tr} S_Q^{bb^\prime T} (x)
C A_1^k S_u^{cc^\prime T} (x) C A_1^\ell \Big] \Big\} \ve 0 \ra_\gamma~,
\eea
where $A_1^1=1$, $A_1^2=t \gamma_5$, $A_2^1=\gamma_5$, $A_2^2=t$,
and $S_q(x)$ and $S_Q(x)$ are the full propagators of the light and heavy
quarks.

The expressions of the correlator functions for the $\Sigma_Q^-$,
$\Sigma_Q^0$, $\Xi_Q^{\prime 0}$, $\Xi_Q^{\prime -}$ and
$\Omega_Q$ can be found by performing the following replacements,
\bea
\label{esbs09}   
\Pi^{\Sigma_Q^-} \es \Pi^{\Sigma_Q^+} (u\to d)~, \nnb \\
\Pi^{\Sigma_Q^0} \es {1\over 2} \left( \Pi^{\Sigma_Q^+} +
\Pi^{\Sigma_Q^-}\right)~, \nnb \\
\Pi^{\Xi_Q^0} \es \Pi^{\Sigma_Q^+} (u\to s)~, \nnb \\
\Pi^{\Xi_Q^-} \es \Pi^{\Sigma_Q^+} (u\to s, s\to d)~.
\eea

The light cone expression of the light quark propagator in external field is
calculated in \cite{Rsbs26} in which it is found that The contributions of
the nonlocal operators such as $\bar{q} G q$, $\bar{q} G^2 q$,
$\bar{q} q \bar{q} q$, are small. Neglecting these operators is justified in
conformal spin expansion \cite{Rsbs27}. Note that in further analysis we
retain only those terms that are linear in quark mass. The expression of the
light quark operator in the presence of external field is given as,
\bea
\label{esbs10}
S_q(x) \es {i \rlap/x\over 2\pi^2 x^4} - {m_q\over 4 \pi^2 x^2} -
{\lla \bar q q \rra\over 12} \left(1 - i {m_q\over 4} \rlap/x \right) -
{x^2\over 192} m_0^2 \lla \bar q q \rra  \left( 1 -
i {m_q\over 6}\rlap/x \right) \nnb \\
\ek i g_s \int_0^1 du \Bigg[{\rlap/x\over 16 \pi^2 x^2} G_{\mu \nu} (ux)
\sigma_{\mu \nu} - {i\over 4 \pi^2 x^2} u x^\mu G_{\mu \nu} (ux) \gamma^\nu \nnb \\
\ek i {m_q\over 32 \pi^2} G_{\mu \nu} \sigma^{\mu
 \nu} \left( \ln {-x^2 \Lambda^2\over 4}  +
 2 \gamma_E \right) \Bigg] + \cdots~,
\eea
where $\Lambda$ is the cut-off energy separating perturbative and
nonperturbative domains, and $\gamma_E$ is the Euler constant.

In calculating the correlation function from the QCD side we also need the
expression for the heavy quarks, whose explicit form in the coordinate space
can be expressed as,
\bea
\label{esbs11}
S_Q(x) \es {m_Q^2 \over 4 \pi^2} \Bigg\{ {K_1(m_Q\sqrt{-x^2}) \over \sqrt{-x^2}} +
i {\rlap/{x} \over -x^2} K_2(m_Q\sqrt{-x^2}) \Bigg\} \nnb \\ 
\ek {g_s \over 16 \pi^2} \int_0^1 du
G_{\mu\nu}(ux) \left[ \left(\sigma^{\mu\nu} \rlap/x + \rlap/x
\sigma^{\mu\nu}\right) {K_1 (m_Q\sqrt{-x^2})\over \sqrt{-x^2}} +
2 \sigma^{\mu\nu} K_0(m_Q\sqrt{-x^2})\right] +\cdots~,
\eea
where $K_i(m_Q\sqrt{-x^2})$ are the modified Bessel functions.
Taking into account the expressions of the light and heavy propagators, the
correlation function given in Eq. (\ref{esbs08}) can be calculated from the
QCD side, from which we observe three different type of contributions: a)
Perturbative contributions, i.e., photon interacts with the quark
propagators perturbatively. Technically this contribution can be calculated
by replacing the one of the free quark operators (the first two terms in
Eqs. (\ref{esbs10}) and (\ref{esbs11})) by,
\bea
\label{esbs12}
S^{free} (x) \to \int d^4y S^{free} (x-y) \not\!\!{A} (y)  S^{free} (x-y)~,
\eea
and the remaining two propagators are the free ones. b) In the case when
photon interacts with the heavy quark perturbatively, the free part must be
removed at least in one of the light quark propagators. c) Nonperturbative
contributions, i.e., photon interacts with the light quark fields at large
distance. This contribution can be calculated by replacing one of the light
quark operators by,
\bea
\label{esbs13}
S_{\alpha\beta}^{ab} \to -{1\over 4} \left( q^a \Gamma_i q^b \right)
\left(\Gamma_i \right)_{\alpha\beta},
\eea
and the remaining quarks constitute the full quark propagators. Here,
$\Gamma_j$ are the full set of Dirac matrices
$\gamma_j=\left\{I,\gamma_5,\gamma_\mu,i
\gamma_\mu\gamma_5,\sigma_{\mu\nu}/\sqrt{2} \right\}$. When Eq.
(\ref{esbs13}) is used in calculation of the nonperturbative contributions,
we see that matrix elements of the form $\la \gamma(q) \ve \bar{q} \Gamma_i
q \ve 0 \ra$ are needed. These matrix elements are defined in terms of the
photon distribution amplitudes in the following way (see \cite{Rsbs28}),

\bea
\label{esbs14}
&&\langle \gamma(q) \vert  \bar q(x) \sigma_{\mu \nu} q(0) \vert  0
\rangle  = -i e_q \bar q q (\varepsilon_\mu q_\nu - \varepsilon_\nu
q_\mu) \int_0^1 du e^{i \bar u qx} \left(\chi \varphi_\gamma(u) +
\frac{x^2}{16} \mathbb{A}  (u) \right) \nnb \\ &&
-\frac{i}{2(qx)}  e_q \qq \left[x_\nu \left(\varepsilon_\mu - q_\mu
\frac{\varepsilon x}{qx}\right) - x_\mu \left(\varepsilon_\nu -
q_\nu \frac{\varepsilon x}{q x}\right) \right] \int_0^1 du e^{i \bar
u q x} h_\gamma(u)
\nnb \\
&&\langle \gamma(q) \vert  \bar q(x) \gamma_\mu q(0) \vert 0 \rangle
= e_q f_{3 \gamma} \left(\varepsilon_\mu - q_\mu \frac{\varepsilon
x}{q x} \right) \int_0^1 du e^{i \bar u q x} \psi^v(u)
\nnb \\
&&\langle \gamma(q) \vert \bar q(x) \gamma_\mu \gamma_5 q(0) \vert 0
\rangle  = - \frac{1}{4} e_q f_{3 \gamma} \epsilon_{\mu \nu \alpha
\beta } \varepsilon^\nu q^\alpha x^\beta \int_0^1 du e^{i \bar u q
x} \psi^a(u)
\nnb \\
&&\langle \gamma(q) | \bar q(x) g_s G_{\mu \nu} (v x) q(0) \vert 0
\rangle = -i e_q \qq \left(\varepsilon_\mu q_\nu - \varepsilon_\nu
q_\mu \right) \int {\cal D}\alpha_i e^{i (\alpha_{\bar q} + v
\alpha_g) q x} {\cal S}(\alpha_i)
\nnb \\
&&\langle \gamma(q) | \bar q(x) g_s \tilde G_{\mu \nu} i \gamma_5 (v
x) q(0) \vert 0 \rangle = -i e_q \qq \left(\varepsilon_\mu q_\nu -
\varepsilon_\nu q_\mu \right) \int {\cal D}\alpha_i e^{i
(\alpha_{\bar q} + v \alpha_g) q x} \tilde {\cal S}(\alpha_i)
\nnb \\
&&\langle \gamma(q) \vert \bar q(x) g_s \tilde G_{\mu \nu}(v x)
\gamma_\alpha \gamma_5 q(0) \vert 0 \rangle = e_q f_{3 \gamma}
q_\alpha (\varepsilon_\mu q_\nu - \varepsilon_\nu q_\mu) \int {\cal
D}\alpha_i e^{i (\alpha_{\bar q} + v \alpha_g) q x} {\cal
A}(\alpha_i)
\nnb \\
&&\langle \gamma(q) \vert \bar q(x) g_s G_{\mu \nu}(v x) i
\gamma_\alpha q(0) \vert 0 \rangle = e_q f_{3 \gamma} q_\alpha
(\varepsilon_\mu q_\nu - \varepsilon_\nu q_\mu) \int {\cal
D}\alpha_i e^{i (\alpha_{\bar q} + v \alpha_g) q x} {\cal
V}(\alpha_i) \nnb \\ && \langle \gamma(q) \vert \bar q(x)
\sigma_{\alpha \beta} g_s G_{\mu \nu}(v x) q(0) \vert 0 \rangle  =
e_q \qq \left\{
        \left[\left(\varepsilon_\mu - q_\mu \frac{\varepsilon x}{q x}\right)\left(g_{\alpha \nu} -
        \frac{1}{qx} (q_\alpha x_\nu + q_\nu x_\alpha)\right) \right. \right. q_\beta
\nnb \\ && -
         \left(\varepsilon_\mu - q_\mu \frac{\varepsilon x}{q x}\right)\left(g_{\beta \nu} -
        \frac{1}{qx} (q_\beta x_\nu + q_\nu x_\beta)\right) q_\alpha
\nnb \\ && -
         \left(\varepsilon_\nu - q_\nu \frac{\varepsilon x}{q x}\right)\left(g_{\alpha \mu} -
        \frac{1}{qx} (q_\alpha x_\mu + q_\mu x_\alpha)\right) q_\beta
\nnb \\ &&+
         \left. \left(\varepsilon_\nu - q_\nu \frac{\varepsilon x}{q.x}\right)\left( g_{\beta \mu} -
        \frac{1}{qx} (q_\beta x_\mu + q_\mu x_\beta)\right) q_\alpha \right]
   \int {\cal D}\alpha_i e^{i (\alpha_{\bar q} + v \alpha_g) qx} {\cal T}_1(\alpha_i)
\nnb \\ &&+
        \left[\left(\varepsilon_\alpha - q_\alpha \frac{\varepsilon x}{qx}\right)
        \left(g_{\mu \beta} - \frac{1}{qx}(q_\mu x_\beta + q_\beta x_\mu)\right) \right. q_\nu
\nnb \\ &&-
         \left(\varepsilon_\alpha - q_\alpha \frac{\varepsilon x}{qx}\right)
        \left(g_{\nu \beta} - \frac{1}{qx}(q_\nu x_\beta + q_\beta x_\nu)\right)  q_\mu
\nnb \\ && -
         \left(\varepsilon_\beta - q_\beta \frac{\varepsilon x}{qx}\right)
        \left(g_{\mu \alpha} - \frac{1}{qx}(q_\mu x_\alpha + q_\alpha x_\mu)\right) q_\nu
\nnb \\ &&+
         \left. \left(\varepsilon_\beta - q_\beta \frac{\varepsilon x}{qx}\right)
        \left(g_{\nu \alpha} - \frac{1}{qx}(q_\nu x_\alpha + q_\alpha x_\nu) \right) q_\mu
        \right]
    \int {\cal D} \alpha_i e^{i (\alpha_{\bar q} + v \alpha_g) qx} {\cal T}_2(\alpha_i)
\nnb \\ &&+
        \frac{1}{qx} (q_\mu x_\nu - q_\nu x_\mu)
        (\varepsilon_\alpha q_\beta - \varepsilon_\beta q_\alpha)
    \int {\cal D} \alpha_i e^{i (\alpha_{\bar q} + v \alpha_g) qx} {\cal T}_3(\alpha_i)
\nnb \\ &&+
        \left. \frac{1}{qx} (q_\alpha x_\beta - q_\beta x_\alpha)
        (\varepsilon_\mu q_\nu - \varepsilon_\nu q_\mu)
    \int {\cal D} \alpha_i e^{i (\alpha_{\bar q} + v \alpha_g) qx} {\cal T}_4(\alpha_i)
                        \right\}~,
\eea
where $\varphi_\gamma(u)$ is the leading twist-2, $\psi^v(u)$,
$\psi^a(u)$, ${\cal A}$ and ${\cal V}$ are the twist-3, and
$h_\gamma(u)$, $\mathbb{A}$, ${\cal T}_i$ ($i=1,~2,~3,~4$) are the
twist-4 photon DAs, and $\chi$ is the magnetic susceptibility.
The measure ${\cal D} \alpha_i$ is defined as
\bea
\label{nolabel05}
\int {\cal D} \alpha_i = \int_0^1 d \alpha_{\bar q} \int_0^1 d
\alpha_q \int_0^1 d \alpha_g \delta(1-\alpha_{\bar
q}-\alpha_q-\alpha_g)~.\nnb
\eea
As has already been noted in determining the magnetic moments of heavy
baryons, we need four equations.
Separating the coefficients of the structures
$(\varepsilon\!\cdot\!p) I$, $(\varepsilon\!\cdot\! p) \rlap/{p}$,
$\rlap/{\varepsilon} \rlap/{p}$ and $\rlap/{\varepsilon}$ 
from the theoretical parts, and equating them
to the corresponding structures in the phenomenological parts, one can
obtain from Eq. (\ref{esbs05}) the sum rules which describe ``positive (negative) parity $\to$
positive (negative) parity" transitions; as well as ``positive (negative)
parity $\to$ negative (positive) parity" transitions. Solving these
equations for the ``negative parity $\to$ negative parity" transitions, the
following sum rules are obtained,
\bea
\label{esbs15}
{\lambda_-^2 \mu \left[2 (m_+ + m_-)^2\right] \over (m_-^2 - p^2)
[m_-^2-(p+q)^2]} =  
- (m_- + m_+) \left( \Pi_1^{th} - m_+ \Pi_2^{th}\right) -
2 m_+ \Pi_3^{th} + 2 \Pi_4^{th}~,
\eea
where $\mu=(f_1^\ast + f_2^\ast)\ve_{q^2=0}$ is the magnetic moment of the
corresponding negative parity baryons.

Performing Borel transformation over the variables $-p^2$ and $-(p+q)^2$ in
order to enhance contributions of the ground states, and suppress the
continuum and higher state contributions; and subtracting the continuum
contributions, we finally obtain the following sum rules for the negative parity
baryons
\bea
\label{esbs16}
\mu = {e^{m_-^2/M^2} \over \lambda_-^2 \left[2 (m_+ + m_-)^2\right] }
\left\{ - (m_- + m_+) \left( \Pi_1^{th} - m_+ \Pi_2^{th}\right) -
2 m_+ \Pi_3^{th} + 2 \Pi_4^{th}\right\}~,
\eea
where we set $M_1^2=M_2^2=2 M^2$.
The expressions of $\Pi_i^{th}$ are quite lengthy, so we do not present
their expressions here.

Our final attempt in this section is the calculation of residues of the
negative parity heavy baryons, which are needed in determination of the
magnetic moments. These residues are determined from an analysis of the
two-point correlator function,
\bea
\label{esbs17}
\Pi^M(q^2)= i \int d^4x e^{iqx} \la 0 \ve \mbox{\rm T} \left\{ \eta_Q(x)
\bar{\eta}_Q(0) \right\} \ve 0 \ra~,
\eea
where the superscript $M$ means mass sum rule.
This correlation function have two structures, namely, $\not\!p$ and unit
matrix $I$, and can be written as,
\bea
\label{esbs18}
\Pi^M(q^2) = \Pi^M_1 \not\!p + \Pi^M_2 I~.
\eea
Saturating (\ref{esbs17}) with positive and negative parity baryons, we get
\bea
\label{esbs19}
\Pi^M (q^2) = {\ve \lambda_- \ve^2 (\not\!p - m_-) \over m_-^2-p^2} +
{\ve \lambda_+ \ve^2 (\not\!p + m_+) \over m_+^2-p^2}~.
\eea
Eliminating the positive parity baryons, and performing Borel transformation
over the variable $-p^2$, we get the following sum rules for the masses and
residues of the negative parity baryons.
\bea
\label{esbs20}
\ve \lambda_- \ve^2 \es {1\over \pi} {e^{m_-^2/M^2} \over m_++m_-} 
\int_{m_b^2}^{s_0} ds e^{-s/M^2} \left[ m_+ \mbox{\rm Im} \Pi^M_1(s) - \mbox{\rm Im}
\Pi^M_2(s) \right]~,\nnb \\
m_-^2 \es {\int_{m_b^2}^{s_0} s ds e^{-s/M^2} \left[  m_+ \mbox{\rm Im}
\Pi^M_1(s) - \mbox{\rm Im} \Pi^M_2(s) \right] \over
\int_{m_b^2}^{s_0} ds e^{-s/M^2} \left[  m_+ \mbox{\rm Im}     
\Pi^M_1(s) - \mbox{\rm Im} \Pi^M_2(s) \right]}~.
\eea
The expressions of $\mbox{\rm Im}\Pi^M_1(s)$ and $\mbox{\rm Im}\Pi^M_2(s)$
for the $\Sigma_b^0$ baryon are
presented in Appendix B.

\section{Numerical analysis}

This section is devoted to the numerical analysis of the sum rules for the
$J^P={1\over 2}^-$ heavy baryons obtained in the previous section.
The main input parameters of the light cone QCD sum rules are the photon
distribution amplitudes (DAs). The photon DAs are obtained in \cite{Rsbs28},
and for completeness we present their expressions in Appendix C. Sum rules
for the magnetic moment, together with the photon DAs, also contain the
following input parameters: quark condensate $\qq$, $m_0^2$ that appears in
determination of the vacuum expectation value of the dimension--5 operator
$\la \bar{q} G q \ra = m_0^2 \qq$, magnetic
susceptibility $\chi$ of quarks, etc. In the present analysis we use $\left[ \uu =
\dd \right]_{\mu=1~GeV} = -(0.243)^3~GeV^3$ \cite{Rsbs29}, 
$\sp \ve_{\mu=1~GeV} = 0.8 \uu\ve_{\mu=1~GeV}$, $m_0^2=(0.8\pm 0.2)~GeV^2$
\cite{Rsbs30}. The values of the magnetic susceptibilities can be found in
numerous works (see for example \cite{Rsbs31,Rsbs32,Rsbs33}). In our
numerical analysis we use $\chi(1~GeV)=-2.85~GeV^2$ obtained in
\cite{Rsbs33}.

Having determined input parameters, in this section we shall proceed with
the analysis of the sum rules for the magnetic moments of the $J^P={1\over
2}^-$ heavy baryons. The sum rules contain three auxiliary parameters:
continuum threshold $s_0$, Borel mass square $M^2$, and $t$ appearing in the
expression of the interpolating current. According to the QCD sum rules
philosophy, any measurable quantity must be independent of these parameters.
For this reason we try to find such regions of these parameters where
magnetic moments are insensitive to their variations. This issue can be
accomplished by the following three-step procedure. First, at fixed values
of $s_0$ and $t$, we try to find region of $M^2$ where magnetic moment is
independent of its variation. The upper bound of $M^2$ is determined by
requiring that the contributions of higher states and continuum constitute,
say, less than 40\% of the contribution coming from the perturbative part.
The lower bound of $M^2$ is obtained by demanding that higher twist
contributions are less than the leading twist contributions. Analysis of our
sum rules leads to the following regions of $M^2$ where magnetic moments are
independent on its variation.
\bea
\label{nolabel06}
&& 2.5~GeV^2 \le M^2 \le 4.0~GeV^2,~\mbox{\rm
for}~\Sigma_c,~\Xi_c^\prime,~\Lambda_c, \Xi_c~, \nnb \\
&& 4.5~GeV^2 \le M^2 \le 7.0~GeV^2,~\mbox{\rm
for}~\Sigma_b,~\Xi_b^\prime,~\Lambda_b, \Xi_b~. \nnb
\eea
Next, we try to the find the domain of variation of the continuum threshold
$s_0$, which is the energy square where the continuum starts. The difference
$\sqrt{s_0}-m$, $m$ being the ground state mass, is the energy needed for the
excitation of the particle to its first excited state, and usually this
difference is varies in the range between $0.3~GeV$ and $0.8~GeV$. In our
analysis we we use the average value $\sqrt{s_0}-m=0.5~GeV$.

As an example, in Fig. (1) we present the dependence of the
magnetic moments of $\Sigma_c^0$ baryon on $M^2$, at
$s_0=12~GeV^2$, and at several fixed values of $t$. It follows from these figures 
that, indeed, we have very good stability of the magnetic moment $\mu$ as
$M^2$ varies in its above-mentioned working region. We also analyze these
dependencies at $s_0=11~GeV^2$ and $s_0=13~GeV^2$; and find out that the discrepancy in the
values of the magnetic moment is about 10\%. In other words, the magnetic
moments of  $\Sigma_c^0$ baryon exhibits the expected
insensitivity to the variations in $s_0$ and $M^2$.

Having decided the working regions of $M^2$ and $s_0$, the third and last
step is to find the working region of the parameter $t$ in which
the predictions for the values magnetic moments of heavy baryons show 
good stability. For this aim, we study the dependence of the magnetic
moment of the  $\Sigma_c^0$ baryon on $\cos\theta$,
where $t=\tan\theta$. We observe from our numerical
analysis that, the magnetic moments of all baryons are independent of the
variation in $\cos\theta$ when it varies in the region $-0.7 \le \cos\theta
\le -0.4$. Our numerical analysis predicts that the magnetic moment of the
$\Sigma_c^0$ baryon is $\mu = (-2.0\pm 0.1) \mu_N$, where $\mu_n$ is the
nuclear magneton. Performing similar analysis we have calculated the
magnetic moments of the other $J^P={1\over 2}^-$ heavy baryons whose 
results are presented in Table 2. We note here that, in many
cases, the naive expectation that the relation between the negative and
positive parity baryons, i.e.,
\bea
\label{nolabel07}
\ve \mu_- \ve = {m_+ \over m_-} \ve \mu_+ \ve~, \nnb
\eea
is violated considerably. This violation can be attributed to the fact that
in our analysis we take into account contributions coming from
positive-to-positive and nondiagonal transitions.

    
\begin{table}[h]

\renewcommand{\arraystretch}{1.3}
\addtolength{\arraycolsep}{-0.5pt}
\small
$$
\begin{array}{|c|c||c|c|}
\hline \hline 
                 & \mu            &                  & \mu           \\ \hline
\Sigma_b^+       &  1.3  \pm 0.3  & \Sigma_c^{++}    &  2.2 \pm 0.2  \\
\Sigma_b^0       &  0.5  \pm 0.05 & \Sigma_c^+       &  0.15\pm 0.02 \\
\Sigma_b^-       & -0.3  \pm 0.1  & \Sigma_c^0       & -2.0 \pm 0.1   \\
\Xi_b^{\prime -} & -0.4  \pm 0.1  & \Xi_c^{\prime 0} & -2.0 \pm 0.2   \\
\Xi_b^{\prime 0} &  0.4  \pm 0.1  & \Xi_c^{\prime +} &  0.15\pm 0.02 \\
\Omega_b^-       & -0.3  \pm 0.1  & \Omega_c^0       & -2.0 \pm 0.2   \\
\Lambda_b^0      & -0.11 \pm 0.02 & \Lambda_c^+      &  1.3 \pm 0.2   \\
\Xi_b^-          & -0.7  \pm 0.1  & \Xi_c^0          &  1.6 \pm 0.2   \\
\Xi_b^0          & -0.12 \pm 0.02 & \Xi_c^+          &  1.2 \pm 0.2   \\
\hline \hline 
\end{array}
$$
\caption{Magnetic moments of the negative parity, spin-1/2 baryons
containing single heavy quark belonging to the
sextet and antitriplet representations, in units of nuclear magneton
$\mu_N$}
\renewcommand{\arraystretch}{1}
\addtolength{\arraycolsep}{-1.0pt}
\end{table}       


In conclusion, we have employed the light cone QCD rules to calculate the
magnetic moments of negative parity baryons containing single heavy quark.
In determination of the magnetic moments of these baryons the contamination
coming from the positive parity baryons, as well as from the transitions
between opposite parity baryons are eliminated by considering different
Lorentz structures. A comparison our results
on the magnetic moments of the negative parity baryons with the predictions
of other approaches, such as quark model, bag model, chiral perturbation
theory, lattice QCD, etc., would be interesting.

\newpage


\section*{Appendix A: Photon distribution amplitudes}
\setcounter{equation}{0}
\setcounter{section}{0}


Explicit forms of the photon DAs
\cite{Rsbs28}:

\bea
\label{nolabel27}
\varphi_\gamma(u) \es 6 u \bar u \Big[ 1 + \varphi_2(\mu)
C_2^{\frac{3}{2}}(u - \bar u) \Big]~,
\nnb \\
\psi^v(u) \es 3 [3 (2 u - 1)^2 -1 ]+\frac{3}{64} (15
w^V_\gamma - 5 w^A_\gamma)
                        [3 - 30 (2 u - 1)^2 + 35 (2 u -1)^4]~,
\nnb \\
\psi^a(u) \es [1- (2 u -1)^2] [ 5 (2 u -1)^2 -1 ]
\frac{5}{2}
    \Bigg(1 + \frac{9}{16} w^V_\gamma - \frac{3}{16} w^A_\gamma
    \Bigg)~,
\nnb \\
{\cal A}(\alpha_i) \es 360 \alpha_q \alpha_{\bar q} \alpha_g^2
        \Bigg[ 1 + w^A_\gamma \frac{1}{2} (7 \alpha_g - 3)\Bigg]~,
\nnb \\
{\cal V}(\alpha_i) \es 540 w^V_\gamma (\alpha_q - \alpha_{\bar q})
\alpha_q \alpha_{\bar q}
                \alpha_g^2~,
\nnb \\
h_\gamma(u) \es - 10 (1 + 2 \kappa^+ ) C_2^{\frac{1}{2}}(u
- \bar u)~,
\nnb \\
\mathbb{A}(u) \es 40 u^2 \bar u^2 (3 \kappa - \kappa^+ +1 ) +
        8 (\zeta_2^+ - 3 \zeta_2) [u \bar u (2 + 13 u \bar u) + 
                2 u^3 (10 -15 u + 6 u^2) \ln(u) \nnb \\ 
\ar 2 \bar u^3 (10 - 15 \bar u + 6 \bar u^2)
        \ln(\bar u) ]~,
\nnb \\
{\cal T}_1(\alpha_i) \es -120 (3 \zeta_2 + \zeta_2^+)(\alpha_{\bar
q} - \alpha_q)
        \alpha_{\bar q} \alpha_q \alpha_g~,
\nnb \\
{\cal T}_2(\alpha_i) \es 30 \alpha_g^2 (\alpha_{\bar q} - \alpha_q)
    [(\kappa - \kappa^+) + (\zeta_1 - \zeta_1^+)(1 - 2\alpha_g) +
    \zeta_2 (3 - 4 \alpha_g)]~,
\nnb \\
{\cal T}_3(\alpha_i) \es - 120 (3 \zeta_2 - \zeta_2^+)(\alpha_{\bar
q} -\alpha_q)
        \alpha_{\bar q} \alpha_q \alpha_g~,
\nnb \\
{\cal T}_4(\alpha_i) \es 30 \alpha_g^2 (\alpha_{\bar q} - \alpha_q)
    [(\kappa + \kappa^+) + (\zeta_1 + \zeta_1^+)(1 - 2\alpha_g) +
    \zeta_2 (3 - 4 \alpha_g)]~,\nnb \\
{\cal S}(\alpha_i) \es 30\alpha_g^2\{(\kappa +
\kappa^+)(1-\alpha_g)+(\zeta_1 + \zeta_1^+)(1 - \alpha_g)(1 -
2\alpha_g)\nnb \\ 
\ar\zeta_2
[3 (\alpha_{\bar q} - \alpha_q)^2-\alpha_g(1 - \alpha_g)]\}~,\nnb \\
\widetilde {\cal S}(\alpha_i) \es-30\alpha_g^2\{(\kappa -
\kappa^+)(1-\alpha_g)+(\zeta_1 - \zeta_1^+)(1 - \alpha_g)(1 -
2\alpha_g)\nnb \\ 
\ar\zeta_2 [3 (\alpha_{\bar q} -
\alpha_q)^2-\alpha_g(1 - \alpha_g)]\}. \nnb
\eea
The parameters entering  the above DA's are borrowed from
\cite{Rsbs28} whose values are $\varphi_2(1~GeV) = 0$, 
$w^V_\gamma = 3.8 \pm 1.8$, $w^A_\gamma = -2.1 \pm 1.0$, 
$\kappa = 0.2$, $\kappa^+ = 0$, $\zeta_1 = 0.4$, $\zeta_2 = 0.3$, 
$\zeta_1^+ = 0$, and $\zeta_2^+ = 0$.


\newpage


\section*{Appendix B} 

In this appendix we give the 
expressions of the invariant amplitudes $\Pi_1^M$ and
$\Pi_2^M$ entering into the mass sum rule for the $\Sigma_b^0$ baryon.
Here in this appendix, and in appendix C the masses of the light quarks are
neglected.

\setcounter{equation}{0}
\setcounter{section}{0}

%
%
\bea
&&\Pi_1^M =
{3\over 256 \pi^4} \Big\{ - m_b^4  M^6  [5 + t (2 + 5 t)] \left[
m_b^4 {\cal I}_5 -
2 m_b^2 {\cal I}_4 +
{\cal I}_3\right] \Big\} \nnb \\
\ar {1\over 192 \pi^4} m_b^4 M^2 \left[\GG  (1 + t + t^2 ) - 18 m_b \pi^2 (-1+t^2)
\left(\dd+\uu\right)\right] {\cal I}_3 \nnb \\
\ar {1\over 3072 \pi^4} m_b^2 M^2 \left[-\GG  (13 + 10 t + 13 t^2 ) + 288 m_b \pi^2
(-1+t^2) \left(\dd+\uu\right)\right] {\cal I}_2 \nnb \\
\ar {e^{-m_b^2/M^2} \over 73728 m_b M^2 \pi^4} \Big\{ - \GG^2 m_b (1 + t)^2 +
768 m_b m_0^2 \dd \uu \pi^4 (-1 + t)^2 \nnb \\
\ek 56 \GG m_0^2 \pi^2 (-1 + t^2) \left(\dd + \uu\right) \Big\} \nnb \\
\ar {1 \over 768 M^2 \pi^2} \GG m_b (-1 + t^2) (\dd + \uu) {\cal I}_1 \nnb \\
\ar {e^{-m_b^2/M^2} \over 18432 M^4  \pi^2} m_b m_0^2
\left[\GG \left(\uu+\dd\right) (-1+t^2) +
384 m_b \dd \uu \pi^2 (-1+t)^2 \right] \nnb \\
\ar {e^{-m_b^2/M^2} \over 1728 M^6} m_b^2 \GG (-1 + t)^2 \dd \uu \nnb \\
\ar {e^{-m_b^2/M^2} \over 1728 M^8} m_b^2 m_0^2 \GG (-1 + t)^2 \dd \uu \nnb \\
\ek {e^{-m_b^2/M^2} \over 3456 M^{10}} m_b^4 m_0^2 \GG (-1 + t)^2 \dd \uu \nnb \\
\ek {e^{-m_b^2/M^2} \over 768 m_b \pi^2} \left[ \GG \left(\uu+\dd\right)
(-1+t^2) + 32 m_b \dd \uu \pi^2 (-1+t)^2 \right] \nnb \\
\ar {1\over 256 \pi^2} m_b \left(\uu+\dd\right) (-1+t^2) \left[
\left(\GG - 13 m_b^2 m_0^2\right) {\cal I}_2 + 6 m_0^2 {\cal I}_1
\right]~, \nnb \\ \nnb \\ \nnb \\
%
%
&&\Pi_2^M =
-{3\over 256 \pi^4} \Big\{ - m_b^3  M^6 (-1+t)^2 \left[          
m_b^4 {\cal I}_4 -                          
2 m_b^2 {\cal I}_3 +                               
{\cal I}_2\right] \Big\} \nnb \\       
\ar {1\over 3072 \pi^4} m_b M^4 \Big\{4 m_b^2 \left[\GG (-1+t)^2 + 72 m_b
\left(\uu+\dd\right) \pi^2 (-1+t^2) \right] {\cal I}_3
- 3 \GG (-1+t)^2 {\cal I}_2 \Big\} \nnb \\
\ek {7 e^{-m_b^2/M^2} \over 256 \pi^2} m_0^2 M^2
\left(\uu+\dd\right) (-1 + t^2) \nnb \\
\ar {1\over 1024 \pi^4} m_b M^2 \Big\{
m_b \left[ 3 m_b \GG (-1+t)^2 + 4 m_0^2 \left(\uu+\dd\right) \pi^2 (-1+t^2)
\right] {\cal I}_2 -
2 \GG (-1+t)^2 {\cal I}_1 \Big\} \nnb \\
\ek {e^{-m_b^2/M^2} \over 73728 M^2 \pi^4} m_b 
\left[ \GG^2 (-1 + t)^2 +
1536 m_0^2 \dd \uu \pi^4 (3 + 2 t + 3 t^2) \right] \nnb \\
\ar {e^{-m_b^2/M^2} \over 18432 M^4  \pi^2} m_b 
\left[ -11 m_b m_0^2 \GG \left(\uu+\dd\right) (-1+t^2)\right. \nnb \\
\ek \left. 32 \left(\GG - 12 m_0^2 m_b^2 \right) \dd \uu (5 + 2 t + 5 t^2) 
\right] \nnb \\
\ar {e^{-m_b^2/M^2} \over 1728 M^6} m_b \left( m_b^2 - 3 m_0^2 \right)
\GG \dd \uu (5 + 2 t + 5 t^2) \nnb \\
\ar {e^{-m_b^2/M^2} \over 576 M^8} m_b^3 m_0^2 \GG \dd \uu (5 + 2 t + 5 t^2)
\nnb \\
\ek {e^{-m_b^2/M^2} \over 3456 M^{10}} m_b^5 m_0^2 \GG \dd \uu (5 + 2 t + 5
t^2) \nnb \\
\ar {e^{-m_b^2/M^2} \over 36864 m_b \pi^4} \left[ \GG^2 (-1+t)^2
- 1536 m_b^2 \dd \uu \pi^4 (5 + 2 t + 5 t^2) \right. \nnb \\
\ar \left.96 m_b \GG \left(\uu+\dd\right) \pi^2 (-1+t^2) \right]~,
\eea
where
\bea
{\cal I}_n = \int_{m_b^2}^{\infty} ds\, {e^{-s/M^2} \over s^n}~.\nnb
\eea



\newpage

\section*{Appendix C}
In this Appendix we present the expressions of the invariant functions
$\Pi_i$ appearing in
the sum rules for the magnetic moment of $\Sigma_b^0$ baryon.\\\\

\setcounter{equation}{0}
\setcounter{section}{0}


{\bf 1) Coefficient of the $(\varepsilon\!\cdot\!p) I$ structure}
%
%
\bea
\Pi_1 \es
-{3\over 128 \pi^4} (-1 + t)^2 (e_b + e_d + e_u) m_b^3 M^3 
  \left( {\cal I}_2 - 2 m_b^2 {\cal I}_3 + m_b^4 {\cal I}_4 \right) \nnb \\
\ar {1\over 1536 \pi^4}
(-1 + t) m_b M^4 \Big\{3 \left[(e_d + e_u) (1 - t) \GG +
48 (1 + t) e_b m_b \pi^2 (\dd + \uu)\right] {\cal I}_2 \nnb \\
\ar 4 m_b^2 \left[ \left(e_d + e_u\right) (-1 + t) \GG + 
      72 m_b (1 + t) \pi^2 \left(e_u \dd +e_d \uu\right) \right]
{\cal I}_3 \Big\} \nnb \\
\ek {3\over 16 \pi^2} (-1 + t^2) m_b^4 M^4 \left (e_d \dd + e_u \uu\right)
\widetilde{j}(h_\gamma) {\cal I}_3 \nnb \\
\ar {1\over 16 \pi^2} (-1 + t)^2 (e_d + e_u) f_{3\gamma} m_b^3
   M^2 \left({\cal I}_2 - m_b^2 {\cal I}_3\right) \psi^v(u_0)\nnb \\
\ar {3\over 32 \pi^2} (-1 + t^2) e_b m_b^2 M^2 (\dd + \uu) {\cal I}_1 \nnb \\
\ar {1\over 768 \pi^2} m_b M^2 {\cal I}_2 \Big\{-3 (-1 + t^2) m_b \left[\dd (7 e_b - 2 e_u) m_0^2 + 
      24 \dd e_b m_b^2 \right. \nnb \\
\ar \left. (7 e_b - 2 e_d) m_0^2 \uu + 24 e_b m_b^2
\uu \right] + 
    2 (-1 + t)^2 (e_d + e_u) f_{3\gamma} \GG \psi^v(u_0)\Big\} \nnb \\
\ek {e^{-m_b^2/M^2}\over 2304 m_b \pi^2}
(-1 + t) M^2 \Big\{9 (1 + t) m_0^2 m_b \left(7 \dd e_b + 12 e_u \dd + 7 e_b \uu + 
      12 e_d \uu\right) \nnb \\
\ar 2 f_{3\gamma} \left[(e_d + e_u) (-1 + t) \GG 
+ 96 (1 + t) m_b \pi^2 \left(e_d \uu + e_u \dd\right)\right] 
     \psi^v(u_0)\Big\} \nnb \\
\ar {e^{-m_b^2/M^2}\over 48 M^2}
(-1 + t^2) f_{3\gamma} m_0^2 m_b^2 \left(e_u \dd + e_d \uu\right)
\psi^v(u_0) \nnb\\
\ar {e^{-m_b^2/M^2}\over 13824 M^4 \pi^2}
(-1 + t^2) \GG m_b^2 \left(e_u \dd + e_d \uu\right) 
  \left[9 m_0^2 + 16 f_{3\gamma} \pi^2 \psi^v(u_0)\right] \nnb \\
\ar {e^{-m_b^2/M^2}\over 1728 M^6}
(-1 + t^2) f_{3\gamma} \GG m_0^2 m_b^2 \left(e_u \dd + e_d \uu\right) \psi^v(u_0) \nnb \\
\ek {e^{-m_b^2/M^2}\over 3456 M^8}  
(-1 + t^2) f_{3\gamma} \GG m_0^2 m_b^4 \left(e_u \dd + e_d \uu\right) \psi^v(u_0) \nnb \\
\ar {1\over 768 \pi^2}
(-1 + t^2) \left[-2 \GG \left(e_u \dd + e_d \uu\right) - 21 e_b m_0^2 m_b^2
\left(\dd + \uu\right) 
     {\cal I}_1\right] \nnb \\
\ar {e^{-m_b^2/M^2}\over 384 \pi^2}
(-1 + t^2) \GG (e_d \dd + e_u \uu) 
   \widetilde{j}(h_\gamma) \nnb \\
\ar {e^{-m_b^2/M^2}\over 96} 
(-1 + t^2) f_{3\gamma} m_0^2 (e_u \dd + e_d \uu) \psi^v(u_0)~. \nnb
\eea
\\\\

{\bf 2) Coefficient of the $(\varepsilon \! \cdot\! p)\!\not\!p$ structure}

\bea
\Pi_2 \es
- {1\over 128 \pi^4} m_b^2 M^6 \Big\{3 (5 + 2 t + 5 t^2) (e_d + e_u) m_b^2 
     \left({\cal I}_3 - 2 m_b^2 {\cal I}_4 + m_b^4 {\cal I}_5\right) \nnb \\
\ar e_b \left[(3 + 2 t + 3 t^2) {\cal I}_2 - 3 (1 + t)^2 m_b^2 {\cal I}_3 - 
      3 (-1 + t)^2 m_b^4 {\cal I}_4 + (3 - 2 t +
3 t^2) m_b^6 {\cal I}_5 \right]\Big\} \nnb \\
\ar {1\over 128 \pi^4}
(3 + 2 t + 3 t^2) e_b m_b^2 M^4 \left(-{\cal I}_1 + 3 m_b^2 {\cal I}_2 - 
   3 m_b^4 {\cal I}_3 + m_b^6 {\cal I}_4\right) \nnb \\
\ar {1\over 1536 \pi^4}
m_b^2 M^2 \Big\{(5 + 2 t + 5 t^2) (e_d + e_u) \GG \nnb \\
\ar 288 (-1 + t^2) m_b \pi^2 \left[ e_u \dd + e_d \uu
+ e_b \left(\dd + \uu\right)\right]\Big\}
\left({\cal I}_2 - m_b^2 {\cal I}_3\right) \nnb \\
\ek {e^{-m_b^2/M^2}\over 768 m_b M^2 \pi^2}
(-1 + t^2) \GG m_0^2 \left(e_u \dd + e_d \uu\right) \nnb \\
\ar {e^{-m_b^2/M^2}\over 1536 M^4 \pi^2}
(-1 + t^2) \GG m_0^2 m_b \left(e_u \dd + e_d \uu\right) \nnb \\
\ek {e^{-m_b^2/M^2}\over 384 m_b \pi^2}
(-1 + t^2) \GG \left(e_u \dd + e_d\uu\right) \nnb \\
\ar {1\over 64 \pi^2}
3 (-1 + t^2) m_0^2 m_b \left (e_u \dd + e_d \uu\right) {\cal I}_1 \nnb \\
\ar {1\over 128 \pi^2} (-1 + t^2) m_b \Big\{\left(e_u \dd + e_d \uu\right) \GG \nnb \\
\ek m_0^2 m_b^2 \left[7 e_b \left(\dd+\uu\right) + 12 \left(e_u \dd + 
 e_d \uu \right) \right] \Big\} {\cal I}_2~.\nnb
\eea
\\\\

{\bf 3) Coefficient of the $\not\!p\!\!\not\!\varepsilon$ structure}

\bea
\Pi_3\es
{1\over 128 \pi^2}
m_b^2 M^4 \Big\{(-1 + t^2) \left(e_d \dd + e_u \uu\right) 
    \Big[\Big({\cal I}_2 - 2 m_b^2 {\cal I}_3\Big) \Big(i_1({\cal S})
+ i_1(\widetilde{\cal S},1) \nnb \\
\ar i_1({\cal T}_2,1) - 
       i_1({\cal T}_4,1) - 2 i_1({\cal T}_2,v) + 2 i_1({\cal T}_4,v)\Big) + 
     12 m_b^2 {\cal I}_3 \widetilde{j}(h_\gamma)\Big] \nnb \\
\ar (-1 + t)^2 (e_d + e_u) f_{3\gamma} m_b 
    \left(-{\cal I}_2 + m_b^2 {\cal I}_3\right) \psi^{a\prime}(u_0) \Big\} \nnb \\
\ar {e^{-m_b^2/M^2}\over 9216 m_b \pi^2}
(-1 + t)^2 (e_d + e_u) f_{3\gamma} \GG M^2 \psi^{a\prime}(u_0) \nnb \\
\ar {e^{-m_b^2/M^2}\over 96}
(-1 + t^2) f_{3\gamma} M^2 \left(e_u \dd +
e_d \uu\right) \psi^{a\prime}(u_0) \nnb \\
\ek {1\over 3072 \pi^2}
(-1 + t)^2 (e_d + e_u) f_{3\gamma} \GG m_b M^2 {\cal I}_2 
\psi^{a\prime}(u_0) \nnb \\
\ek {e^{-m_b^2/M^2}\over 384 M^2}
(-1 + t^2) f_{3\gamma} m_0^2 m_b^2 \left(e_u \dd + e_d \uu\right) 
   \psi^{a\prime}(u_0) \nnb \\
\ek {e^{-m_b^2/M^2}\over 6912 M^4}
(-1 + t^2) f_{3\gamma} \GG m_b^2 \left(e_u \dd + e_d \uu\right) 
   \psi^{a\prime}(u_0) \nnb \\
\ek {e^{-m_b^2/M^2}\over 13824 M^6}
(-1 + t^2) f_{3\gamma} \GG m_0^2 m_b^2 \left(e_u \dd + e_d \uu\right) 
   \psi^{a\prime}(u_0) \nnb \\
\ar {e^{-m_b^2/M^2}\over 27648 M^8} 
(-1 + t^2) f_{3\gamma} \GG m_0^2 m_b^4 \left(e_u \dd + e_d \uu\right) 
  \psi^{a\prime}(u_0) \nnb \\
\ar {e^{-m_b^2/M^2}\over 9216 \pi^2} 
(-1 + t^2) \Big\{\GG \left(e_d \dd + e_u \uu\right) \Big[i_1({\cal S}) +
i_1(\widetilde{\cal S},1) + i_1({\cal T}_2,1) - 
     i_1({\cal T}_4,1) \nnb \\
\ek 2 i_1({\cal T}_2,v) + 2 i_1({\cal T}_4,v) -
12 \widetilde{j}(h_\gamma)\Big] - 
4 f_{3\gamma} m_0^2 \pi^2 \left(e_u \dd + e_d \uu\right)
\psi^{a\prime}(u_0)\Big\}~.
\eea
\\\\

{\bf 4) Coefficient of the $\not\!\varepsilon$ structure}

\bea
\Pi_4\es
{1\over 256 \pi^4} 
m_b^2 M^8 \Big\{3 (e_d + e_u) m_b^2 \Big[(1 + t)^2 {\cal I}_3 + 
     4 (1 + t^2) m_b^2 {\cal I}_4 - (5 + 2 t + 5 t^2) m_b^4 {\cal I}_5\Big] \nnb \\ 
\ek e_b \Big[(3 + 2 t + 3 t^2) {\cal I}_2 + 3 (1 + t)^2 m_b^2 {\cal I}_3 - 
     3 (3 + 2 t + 3 t^2) m_b^4 {\cal I}_4 +
(3 - 2 t + 3 t^2) m_b^6 {\cal I}_5\Big]\Big\} \nnb \\
\ar {1\over 256 \pi^4}
e_b m_b^2 M^6 \Big[4 t {\cal I}_1 + 3 m_b^2 {\cal I}_2 - 6 t m_b^2 {\cal I}_2 + 
    3 t^2 m_b^2 {\cal I}_2 - 6 m_b^4 {\cal I}_3 - 6 t^2 m_b^4 {\cal I}_3 + 
    (3 + 2 t + 3 t^2) m_b^6 {\cal I}_4 \Big]\nnb \\
\ar {1\over 128 \pi^2}
(e_d + e_u) f_{3\gamma} m_b^2 M^6 \Big[(1 + t)^2 {\cal I}_2 - 
    2 (1 + 4 t + t^2) m_b^2 {\cal I}_3\Big] i_2({\cal A},v) \nnb \\
\ar {1\over 128 \pi^2}
(e_d + e_u) f_{3\gamma} m_b^2 M^6 \Big[(1 + t)^2 {\cal I}_2 - 
    4 (1 + t + t^2) m_b^2 {\cal I}_3\Big] i_2({\cal V},v) \nnb \\
\ar {1\over 32 \pi^2}
(3 + 2 t + 3 t^2) (e_d + e_u) f_{3\gamma} m_b^4 M^6 {\cal I}_3 \psi^v(u_0)) \nnb \\
\ar {1\over 64 \pi^2}
(-1 + t^2) m_b^3 M^6 \left(e_d \dd + e_u \uu \right) \chi 
   (-{\cal I}_2 + m_b^2 {\cal I}_3) \varphi_\gamma^\prime (u_0) \nnb \\
\ek {1\over 128 \pi^2}  
 (3 + 2 t + 3 t^2) (e_d + e_u) f_{3\gamma} m_b^4 M^6 {\cal I}_3] 
   \psi^{a\prime}(u_0) \nnb \\
\ar {e^{-m_b^2/M^2}\over 4608 m_b \pi^2} 
(-1 + t^2) \GG M^4 \left(e_d \dd + e_u \uu \right) \chi \varphi_\gamma^\prime(u_0) \nnb \\
\ar {1 \over 3072 \pi^4}
m_b^4 M^4 \Big\{-(5 + 2 t + 5 t^2) (e_d + e_u) \GG \nnb \\
\ek 288 (-1 + t^2) m_b \pi^2 \left[\dd (e_b + e_u) + (e_b + e_d) \uu \right] \Big\} {\cal I}_3 \nnb \\
\ar {1 \over 128 \pi^2}
(-1 + t^2) m_b M^4 \left(e_d \dd + e_u \uu\right) 
   \Big[i_1({\cal S},1) - i_1(\widetilde{\cal S},1) + i_1({\cal T}_1,1) \nnb \\
\ek i_1({\cal T}_2,1) + i_1({\cal T}_3,1) - i_1({\cal T}_4,1)\Big]  {\cal I}_1 \nnb \\
\ar {1 \over 1536 \pi^4}
 m_b M^4 \Big\{
   m_b \Big[(1 + t^2) (e_d + e_u) \GG + 
      144 (-1 + t^2) e_b m_b \pi^2 \left(\dd + \uu\right)\Big] \nnb \\
\ek (-1 + t^2) \pi^2 
     \left(e_d \dd + e_u \uu\right) \Big[6 m_b^2 \Big(2 i_1({\cal S},1) - 2 (i_1(\widetilde{\cal S},1) - 2 i_1({\cal T}_1,1) + 
          i_1({\cal T}_2,1) + i_1({\cal T}_4,1) \nnb \\
\ek 6 i_1({\cal S},v) - 2 i_1({\cal T}_3,v) + 
          2 i_1({\cal T}_4,v)) - \mathbb{A}^\prime (u_0)\Big) + 
      \GG \chi \varphi_\gamma^\prime(u_0)\Big]\Big\} {\cal I}_2 \nnb \\
\ek {e^{-m_b^2/M^2}\over 4608 m_b \pi^2}
(-1 + t^2) \GG M^2 \left(e_d \dd + e_u \uu\right) \Big[i_1({\cal T}_1,1) - i_1({\cal T}_3,1) \nnb \\ 
\ar 6 i_1({\cal S},v) + 2 i_1({\cal T}_3,v) - 2 i_1({\cal T}_4,v)\Big] \nnb \\
\ar {e^{-m_b^2/M^2}\over 9216 \pi^2}
(e_d + e_u) f_{3\gamma} \GG M^2 \Big[(1 + 6 t + t^2) i_2({\cal A},v) + 
   (3 + 2 t + 3 t^2) i_2({\cal V},v)\Big] \nnb \\
\ar {e^{-m_b^2/M^2}\over 2304 \pi^2}
f_{3\gamma} M^2 \Big[-(3 + 2 t + 3 t^2) (e_d + e_u) \GG - 
    288 (-1 + t^2)  m_b \pi^2 \left(e_u \dd + e_d \uu\right) \Big] 
   \psi^v(u_0) \nnb \\
\ar {1\over 256 \pi^2}
(-1 + t^2) m_b M^2 \Big\{e_u \dd \GG - \dd (7 e_b + 12 e_u) m_0^2 m_b^2 \nnb \\
\ar \Big[e_d \GG - (7 e_b + 12 e_d) m_0^2 m_b^2\Big] \uu \Big] {\cal I}_2 \nnb \\
\ar {e^{-m_b^2/M^2}\over 9216 m_b \pi^2}
 (-1 + t^2) M^2 \Big[-36 \left(\GG - 6 m_0^2 m_b^2\right) \left( e_u \dd + e_d \uu\right) \nnb \\ 
\ar \GG \left(e_d \dd + e_u \uu\right) \mathbb{A}^\prime (u_0)\Big] \nnb \\
\ar {e^{-m_b^2/M^2}\over 9216 \pi^2} 
 f_{3\gamma} M^2 \Big[(3 + 2 t + 3 t^2) (e_d + e_u) \GG +
     288 (-1 + t^2) m_b \pi^2 \left(e_u \dd + e_d \uu\right)\Big] \psi^{a\prime}(u_0) \nnb \\
\ar {e^{-m_b^2/M^2}\over 768 M^2 \pi^2}
(-1 + t^2) m_b \left(e_u \dd + e_d \uu\right) 
  \Big[\GG m_0^2 + \pi^2 f_{3\gamma} \left(\GG - 6 m_0^2 m_b^2\right)\nnb \\
\cp \left(-4 \psi^v(u_0) + \psi^{a\prime}(u_0)\right)\Big]] \nnb \\
\ar {e^{-m_b^2/M^2}\over 9216 M^4 \pi^2}
(-1 + t^2) \GG m_b \left(e_u \dd + e_d \uu\right) \Big[-3 m_0^2 m_b^2 \nnb \\
\ek 2 \pi^2 f_{3\gamma} (3 m_0^2 - 2 m_b^2) \left(4 \psi^v(u_0) - 
     \psi^{a\prime}(u_0)\right)\Big] \nnb \\
\ar {e^{-m_b^2/M^2}\over 1536 M^6}
(-1 + t^2) f_{3\gamma} \GG m_0^2 m_b^3 \left(e_u \dd + e_d \uu\right) 
  \left(4 \psi^v(u_0) - \psi^{a\prime}(u_0)\right) \nnb \\
\ek {e^{-m_b^2/M^2}\over 9216 M^8}
(-1 + t^2) f_{3\gamma} \GG m_0^2 m_b^5 \left(e_u \dd + e_d \uu\right)
   \left(4 \psi^v(u_0) - \psi^{a\prime}(u_0)\right) \nnb \\
\ar {e^{-m_b^2/M^2}\over 18432 \pi^2}
(-1 + t^2) \GG m_b \left(\dd e_d + e_u \uu\right) \Big[2 i_1({\cal T}_1,1) - 2 i_1({\cal T}_3,1) + 
    12 i_1({\cal S},v) \nnb \\
\ar 4 i_1({\cal T}_3,v) - 4 i_1({\cal T}_4,v) - 
    \mathbb{A}^\prime(u_0)\Big] \nnb \\
\ek {e^{-m_b^2/M^2}\over 1536 m_b \pi^2} 
 (-1 + t^2) \left(e_u \dd + e_d \uu\right) \Big[\GG (m_0^2 - 2 m_b^2) + 
    12 f_{3\gamma} m_0^2 m_b^2 \pi^2 \left(4 \psi^v(u_0) -
\psi^{a\prime}(u_0)\right)\Big]~. \nnb
\eea

The functions $i_n~(n=1,2)$, $\widetilde{j}_1(f(u))$, and ${\cal I}_n$
are defined as:
\bea
\label{nolabel}
i_1(\phi,f(v)) \es \int {\cal D}\alpha_i \int_0^1 dv 
\phi(\alpha_{\bar{q}},\alpha_q,\alpha_g) f(v) \delta^\prime(k-u_0)~, \nnb \\
i_2(\phi,f(v)) \es \int {\cal D}\alpha_i \int_0^1 dv 
\phi(\alpha_{\bar{q}},\alpha_q,\alpha_g) f(v) \delta^{\prime\prime}(k-u_0)~, \nnb \\
\widetilde{j}(f(u)) \es \int_{u_0}^1 du f(u)~, \nnb \\
{\cal I}_n \es \int_{m_b^2}^{\infty} ds\, {e^{-s/M^2} \over s^n}~,\nnb
\eea
where 
\bea
k = \alpha_q + \alpha_g \bar{v}~,~~~~~u_0={M_1^2 \over M_1^2
+M_2^2}~,~~~~~M^2={M_1^2 M_2^2 \over M_1^2 +M_2^2}~, \mbox{and}~
\bar{v}=1-v~.\nnb
\eea




\newpage

\section*{Figure captions}
{\bf Fig. (1)} The dependence of the magnetic moment of the $\Sigma_c^0$ 
baryon on $M^2$, at several fixed values of $t$, and at
$s_0=12.0~GeV^2$.\\\\
{\bf Fig. (2)} The dependence of the magnetic moment of the $\Sigma_c^0$
baryon on $\cos\theta$, at several fixed values of $M^2$, and at
$s_0=12.0~GeV^2$.

\newpage

\begin{figure}
\vskip 3. cm
    \includegraphics{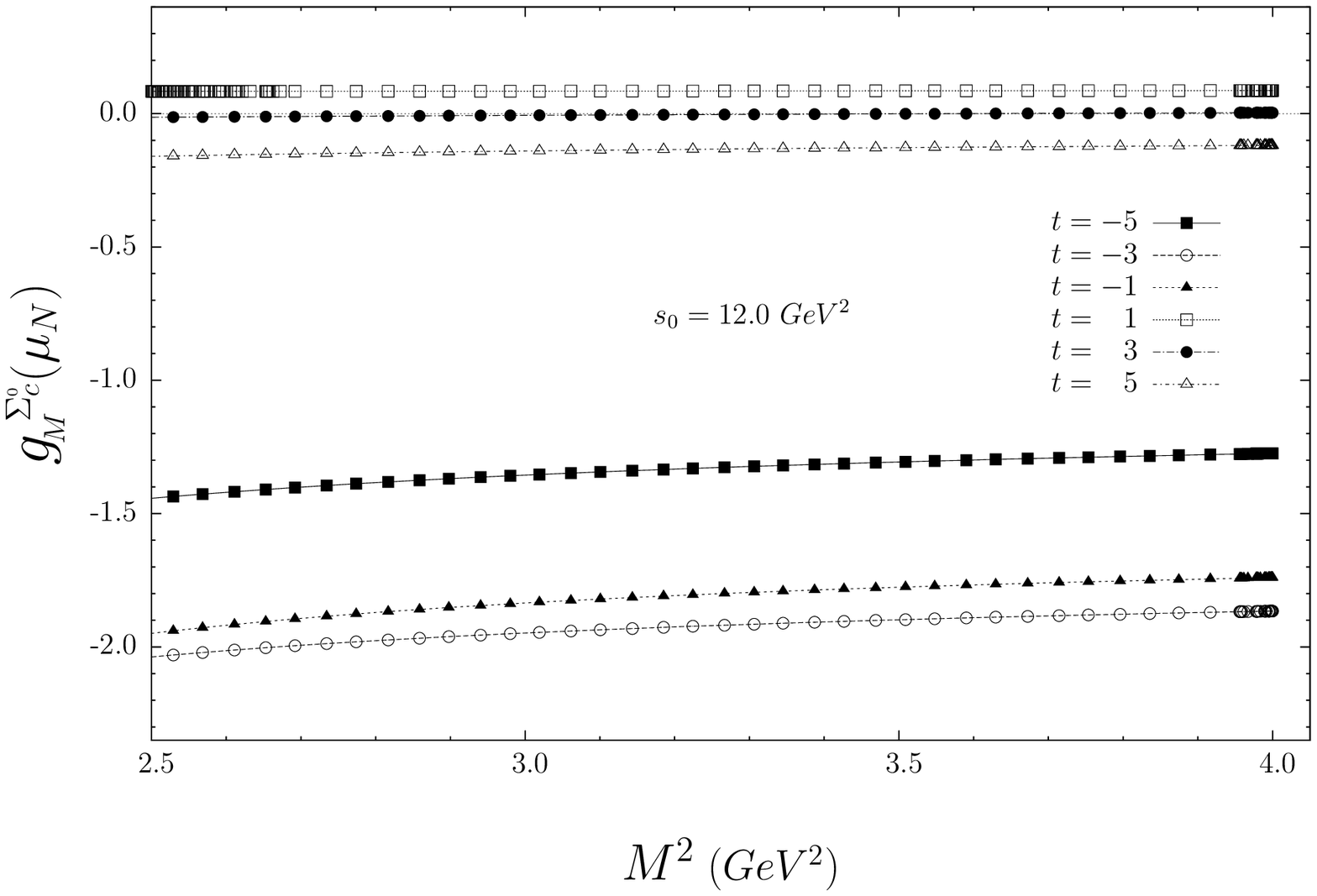}
\vskip 7.0cm
\caption{}
\end{figure}

\begin{figure}
\vskip 3. cm
    \includegraphics{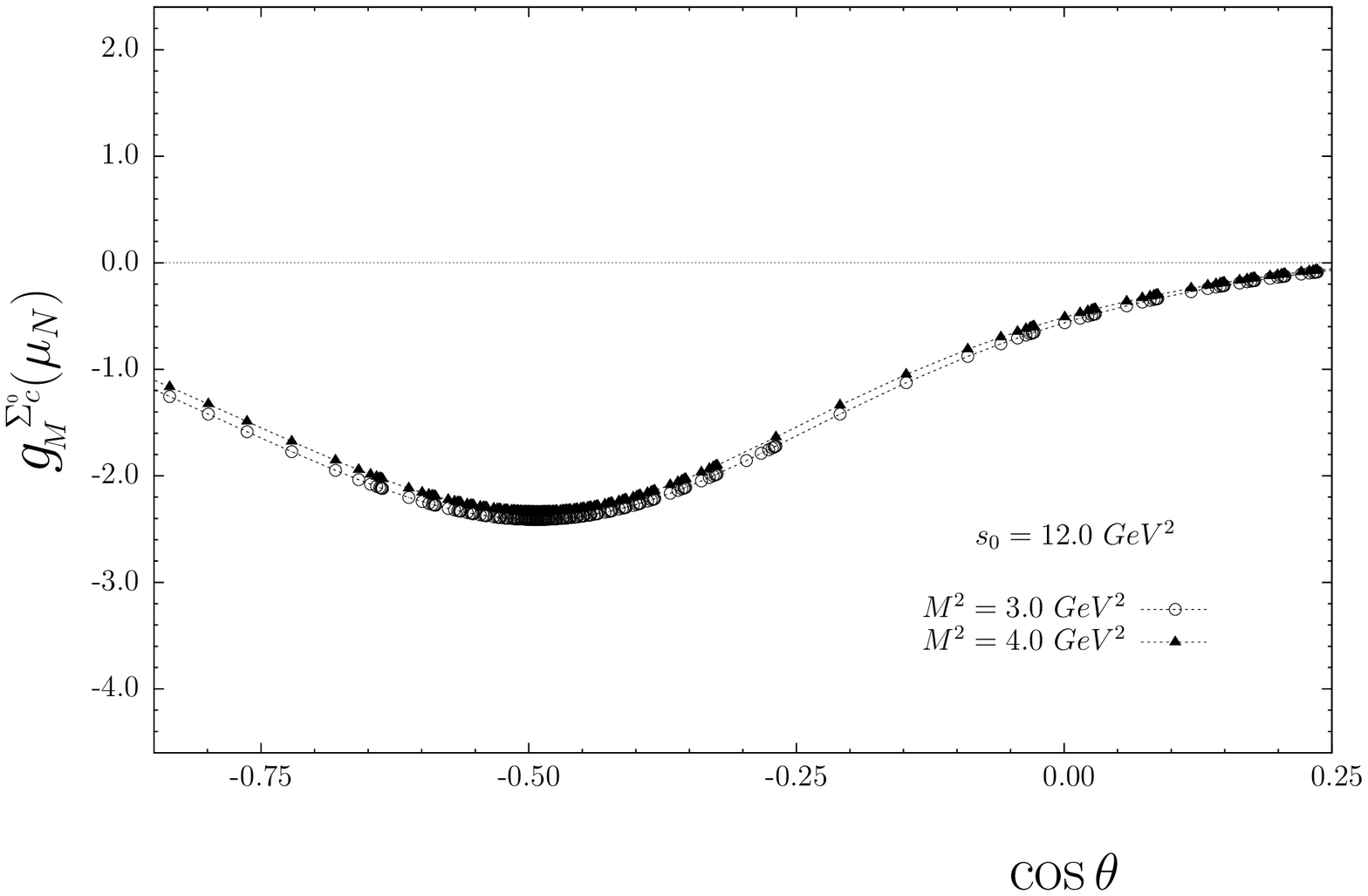}
\vskip 7.0cm
\caption{}
\end{figure}

\end{document}